\documentclass{aastex}
\usepackage{spr-astr-addons}
\usepackage{url}

\RequirePackage{color}
\def\pacce{{\sc pacce} }
\def\w{$W_{\lambda}$}

\begin{document}

\title{{\sc pacce}: Perl Algorithm to Compute Continuum and Equivalent Widths}
\slugcomment{Not to appear in Nonlearned J., 45.}
\shorttitle{{\sc pacce}}
\shortauthors{Riffel R. \& Vale T. B.}

\author{Rog\'erio Riffel\altaffilmark{1}}
\affil{Departamento de Astronomia, Universidade Federal do Rio Grande do Sul. 
              Av. Bento Gon\c calves 9500, Porto Alegre, RS, Brazil.}
\email{riffel@ufrgs.br}
\and
\author{Tib\'erio Borges Vale\altaffilmark{1}}
\affil{Departamento de Astronomia, Universidade Federal do Rio Grande do Sul. 
              Av. Bento Gon\c calves 9500, Porto Alegre, RS, Brazil.}
\email{tiberio.vale@ufrgs.br}


\begin{abstract}

We present Perl Algorithm to Compute continuum and Equivalent Widths ({\sc pacce}). We describe the methods used in the computations and
the requirements for its usage.
We compare the measurements made with \pacce\  and ``manual" ones made using {\sc iraf} {\it splot} task. These tests 
show that for SSP models the equivalent widths strengths are very similar (differences $\lesssim$\,0.2\,\AA) for both measurements. 
In real stellar spectra, the correlation between both values is still very good, but with differences of up to 0.5\,\AA.
\pacce\ is also able to determine mean continuum and continuum at line center values, which are helpful in stellar
population studies. In addition, it is also able to compute the uncertainties in the equivalent widths using photon statistics. 
The code is made available for the community through the web at http://www.if.ufrgs.br/$\sim$riffel/software.html.
 
\end{abstract}



\section{Introduction}

The equivalent widths (\w) of absorption lines, observed in the spectrum of 
astronomical sources can be seen as compressed, but 
highly informative, representation of the whole spectrum. For example, the \w\ of the absorption lines observed 
in galaxies spectra reveals insights about their stellar populations, like the ages
and metallicities of the stars which dominates the light of the host galaxy \citep[e.g.][]{bica88,
schmitt96,rickes04,rembold07,krabbe07,krabbe08,riffel08}. Regarding the spectrum of a star, it is 
possible to make use of the \w\ of absorption lines to determine directly the fundamental atmospheric 
parameters such as: surface gravity ($\log g$), effective temperature ($T_{\rm eff}$) and
the chemical abundances of many elements \citep[see for example][]{gonz96,felt01}.

However, the price to be paid when using the powerful informations contained in the \w\ is the long time
needed to make a reliable measurement of this observables. Commonly, the \w\ are measured using 
interactive routines like {\it splot} provided by the IRAF\footnote{IRAF is distributed 
by National Optical Astronomy Observatories, operated by the Association of Universities for 
Research in Astronomy, Inc., under contract with the National Science Foundation, U.S.A.} team or with
independent codes like LINER \citep{pow93}. Both softwares are ``hand operated", which means the user need to 
look for the line limits and continuum points in the spectrum. The next step, is to mark them
``manually". This procedure is very time-consuming and introduce many uncertainties which are propagated to
a posterior analyses of the quantities involving the \w\ measurements.

With the growing of spectral surveys (e.g. Sloan Digital Sky Survey), it becomes necessary to 
accelerate and automate some process such as the analysis of stellar populations of galaxies, as well as the determination 
of fundamental atmospheric parameters of individual stars. In order
to help in such task we present in this paper a new automatic\footnote{Also interactive, as 
the input parameters are easy to be changed.} code: {\it PACCE: Perl Algorithm to Compute Continuum and 
Equivalent Widths}. This software, written in PERL, can be used to compute the \w\ of absorption lines as 
well as to determine continuum points, being very helpful to perform stellar population synthesis following, 
for example the method developed by \citet{bica88} and \citet{schmitt96}.

This paper is structured as follows: In Sec.~\ref{code} we describe the system requirements as well as the 
numerical procedures behind the code. The input parameters and the outputs of the code are discussed 
in Sec.~\ref{run}. A comparison with the measures made whit \pacce\ and ``hand-made" measurements are presented
in Sec.~\ref{test}. The final remarks are made in Sec.~\ref{final}.

\section{The code}\label{code}

The idea behind \pacce is to reproduce the ``manual" procedure used to measure the \w\ of absorption 
lines in a spectrum, as well as to measure mean continuum fluxes in defined regions and compute the continuum
value at line center. In addition, using the same inputs, it does exactly reproduce the measured values being ``user" independent 
(e.g. the uncertainties introduced by the user in ``hand operated" procedures are removed), and thus allowing for a better comparison 
between \w\ measured by different users.

\pacce was written in Perl, allowing anyone to use it without having any problems with 
software licenses. It is freely distributed under GNU General Public 
License\footnote{http://www.gnu.org/licenses/gpl.html}(GLP). All the 
libraries used in the code are also free and distributed under GLP license.
\pacce\ source code can be freely downloaded from http://www.if.ufrgs.br/$\sim$riffel/software.html. 
All requirements to run \pacce can easily be installed in any linux 
machine ({e.g. {\it apt-get, synaptic} or {\it yum}}), they are:

\begin{itemize}
 \item Perl \\(http://www.perl.org);
 \item Perl's "Math::Derivative" package \\(http://search.cpan.org);
 \item Perl's "Math::Spline" package \\(http://search.cpan.org);
 \item Gnuplot \\(www.gnuplot.info).
\end{itemize}

In addition, we call the attention to the fact that Perl is installed as default in any linux flavor and \pacce\ can 
easily be converted to run under Microsoft Windows system (without plots).

\begin{figure}[h]
\centering
\includegraphics[scale=0.4]{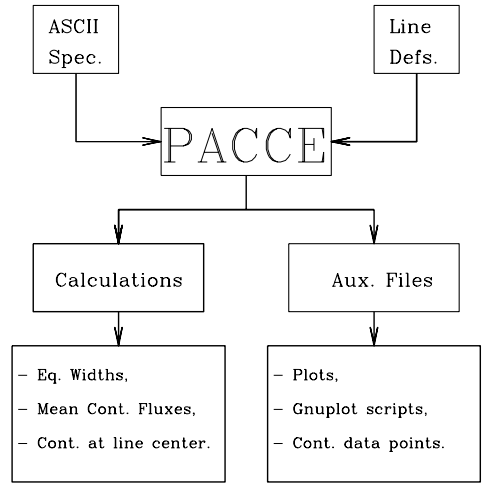}
 \caption{Fluxogram describing the use of \pacce.} 
 \label{fluxog}  
 \end{figure}

\subsection{Calculations: Equivalent Widths and Mean Continuum Fluxes}

\subsubsection{Equivalent Widths}
\begin{figure}[h]
\centering
\includegraphics[scale=0.5]{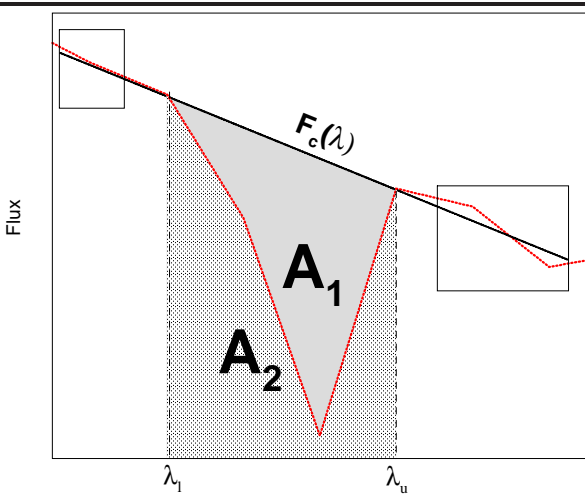}
 \caption{Sketch of the definition of a hypothetical index. Sidebands continuum regions are represented by boxes. Solid 
line is the pseudo-continuum. Vertical lines represent the line limits $\lambda_l$ and $\lambda_u$. Shaded region (A$\rm _2$) 
is the area below the line limits and filled (A$\rm _1$) region is the area of the line. Dotted line (red) represents the spectrum.} 
 \label{ew}  
 \end{figure}

In general, absorption feature indices are composed by measurements of relative flux 
in a central wavelength interval corresponding to the absorption feature considered (line limits $\lambda_l$ and 
$\lambda_u$ in Fig.~ \ref{ew})
and two continuum sidebands passband regions (spectrum ranges - dots - in the boxes of Fig.~ \ref{ew}). Such sidebands provide a 
reference level (pseudo-continuum, solid line Fig.~ \ref{ew}) from which the strength of 
the absorption feature is evaluated \citep[see][for details]{worthey94}. \pacce computes the pseudo-continuum, 
$F_c(\lambda)$, in three ways: (i) a linear regression is computed using all the continuum points 
in the passband regions and a straight line (y=ax+b) is computed using the regression coefficients; 
(ii) a straight line is drawn connecting the mid-points of the flanking passband continuum regions; (iii) 
as a cubic spline\footnote{Such option is useful when measuring \w\ of absorption lines located close to emission lines, and thus, only points - not regions- of continuum free from emission/absorption are available}. The form in which the pseudo-continuum is adjusted is chosen by the user (see Sec.~\ref{run}).

Considering $F_i(\lambda)$ as observed flux per unit wavelength an \w\ is then:
\begin{equation}
\centering
W_{\lambda} =  \displaystyle \int_{\lambda_l}^{\lambda_u} \left(1-\frac{F_i(\lambda)}{F_c(\lambda)}\right)d\lambda
\end{equation}

Considering Fig.~ \ref{ew} one can write the area A$_{\rm 2}$ as:
\begin{equation}
A{\rm _2 =}\displaystyle \int_{\lambda_l}^{\lambda_u}F_i(\lambda)d\lambda
\end{equation}
and, thus, the A${\rm _1}$ is the absorbed flux between $\lambda_l$ and $\lambda_u$. Similarly,
the area below the pseudo-continuum between $\lambda_l$ and $\lambda_u$ is:
\begin{equation}
A{\rm _1} + A{\rm _2 = C =}\displaystyle \int_{\lambda_l}^{\lambda_u}F_c(\lambda)d\lambda,
\end{equation}
thus, the \w\ of the absorption line between $\lambda_l$ and $\lambda_u$ can be written as:
\begin{equation}
W_{\lambda} =  \displaystyle \left(1-\frac{A_2}{C}\right)(\lambda_u-\lambda_l),
\end{equation}
for more details see for example \citet{volmann06}.

In \pacce\ we compute both areas, A$\rm _2$ and C, using the trapezium method. In addition, \pacce\ does 
compute the uncertainties in the equivalent widths, $\sigma(W_{\lambda}$), considering the 
photon noise statistics pixel by pixel. For this purpose we follow \citet{volmann06}, assuming
that the ratio between C and A$\rm _2$ are similar to the normalized ratio 
between these quantities \citep[for details see][]{volmann06}. In the case of linear pseudo-continuum adjustment we 
estimate the signal-to-noise ratio (S/N) as being the ratio between the  square root of the variance
and the mean flux of the points in the bandpass interval (i.e. boxes in Fig~ \ref{ew}). In the 
case of a pseudo-continuum defined with a cubic spline the S/N is calculate in the same 
way as in the linear adjustment, but using points in a region free from emission/absorption lines defined
by the user (see Sec.~\ref{run}).

\subsubsection{Mean Continuum Fluxes}
Besides the \w\, the code also computes the mean continuum fluxes ($\bar{F_{\lambda}}$, last line of Tab.~\ref{inpt})
and the continuum at line center. Such measurements are useful, for example, to perform stellar population studies 
using the technique described by \citet{bica88}. The $\bar{F_{\lambda}}$ point are calculated following the equation:
\begin{equation}
\bar{F_{\lambda}} = \displaystyle \frac{1}{N}\sum_{i=0}^{i=N} F_{\lambda}^{i},
\end{equation}
where $F_{\lambda}^{i}$ is the flux of each $\lambda$ in the defined interval and N is the number of points considered.  The errors are estimated as 
being the square root of the variance. The continuum at the line center
is taken as being the pseudo-continuum flux where $\lambda=\frac{\lambda_l+\lambda_u}{2}$.

\section{Running \pacce}\label{run}

To run \pacce the user needs a ASCII, one dimensional, spectrum and a ASCII input table containing 
the line definitions. The ASCII spectrum is easily created using the {\it wspectext} 
{\sc iraf} task. An example of a input table is shown in Tab.~\ref{inpt}. Note that the continuum points used to determine the line 
pseudo-continuum can be defined into three different ways (linear, mid-point or spline). The fact that such input 
table is very easy to be edited/created, makes \pacce\ also an interactive tool, allowing for fast changes and tests 
in \w\ and $\bar{F_{\lambda}}$ measurements.

\begin{figure}[h]
\centering
\includegraphics[scale=0.65]{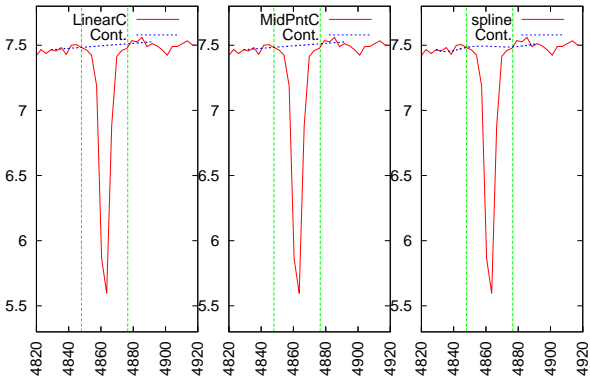}
 \caption{Example of \pacce\ output plots using Tab.~\ref{inpt} index definitions. Mid-point, linear and 
 spline pseudo-continuum adjustments where used in left, center and right plots respectively. The shown region is around H$\beta$ absorption line.} 
 \label{hb}  
 \end{figure}

Besides the calculations, \pacce\ outputs some auxiliary files (Fig.~ \ref{fluxog}). 
These files are the pseudo-continuum data points, the plots showing the regions used in the \w\ calculations (see Fig.~ \ref{hb}), 
as well as the Gnuplot commands, which are stored for future use.

\begin{center}
\begin{table*}
\begin{verbatim}
    #####################################################################################
    #                                                                                   #
    #                   Index definitions table example.                                #
    #                                                                                   #
    #####################################################################################
    #  ID             Line Limits          Left side Cont.     Right side Cont.
       LinearC       (4847.875,4876.625)  [4827.875:4847.875],[4876.625:4891.625] 
       MidPntC       (4847.875,4876.625)  [4827.875:4847.875],[4876.625:4891.625]* 
    # Note the * in the mid-point continuum adjustment.      
    ######################### Spline continuum example ##################################
    #  ID         Line Band Pass       Points for a Spline Cont.              S/N cont.
    spline    (4847.875,4876.625) {4827.875,4833.0,4847.875,4876.625,4891.625}<4880:4890>
    #####################################################################################
    #                                                                                   #
    #                         Intervals for Mean continuum                              #
    #                         see Bica (1986)                                           #
    #####################################################################################
    #
    Cont. |5290:5310,5303:5323,5536:5556,5790:5810,5812:5832,5860:5880,6620:6640|
\end{verbatim}
\caption{Example of input table. A \# indicates that the line is not used in the computations. Note the $*$ in the mid-point continuum adjustment. A full table is available at http://www.if.ufrgs.br/$\sim$riffel/software.html.}
\label{inpt}
\end{table*}
\end{center}

\section{Testing \pacce}\label{test}

In order to test \pacce\ we perform ``hand-made" measurements in a set of \citet{maraston05} 
simple stellar population (SSP) models as well as in a library of observed stellar optical spectra \citep{silva92}.

\begin{figure*}[h]
\centering
\includegraphics[scale=0.80]{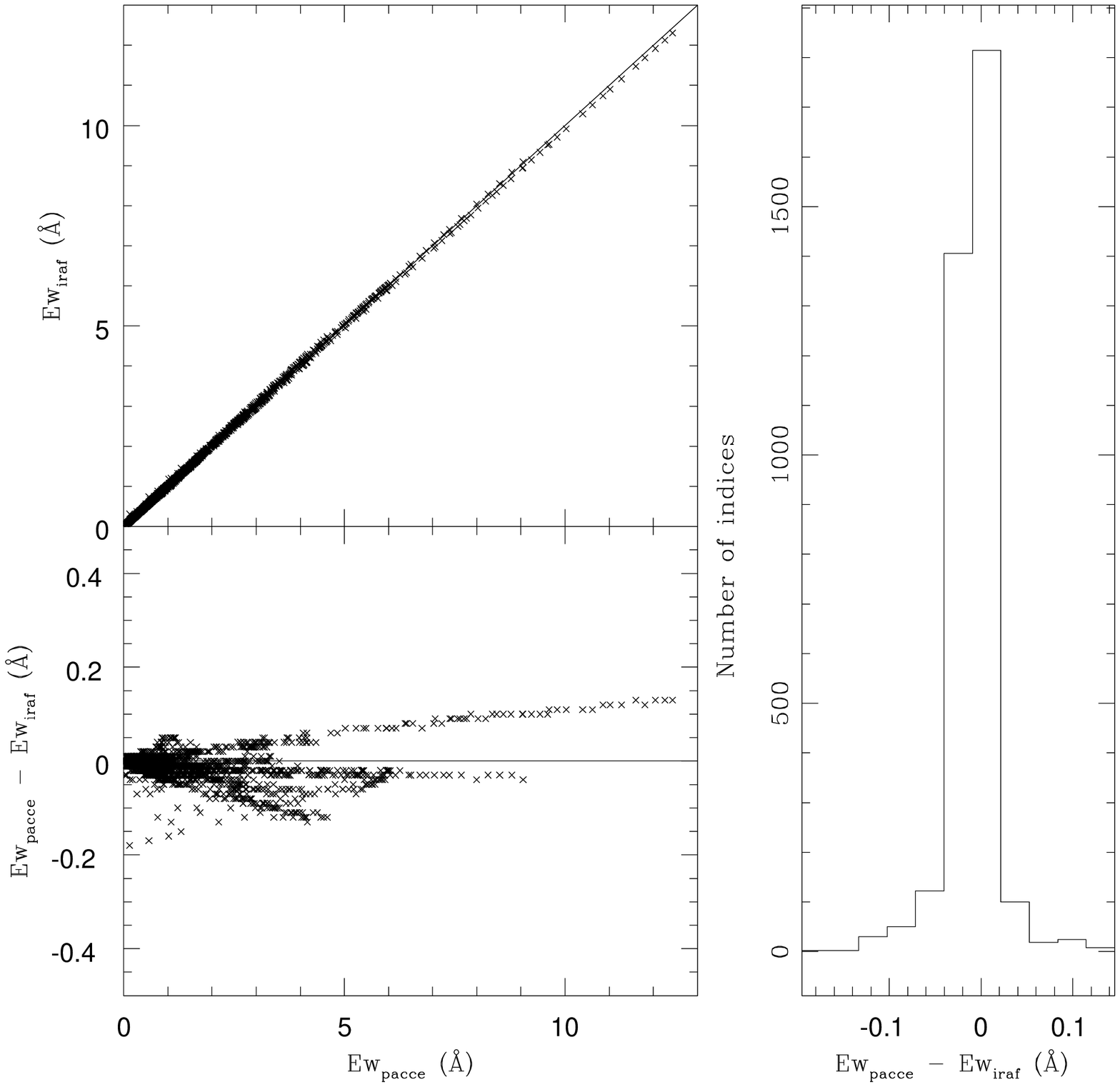}
 \caption{Comparation of \pacce\ and ``hand made" {\sc iraf splot} measurements using \citet{maraston05} 
theoretical models. Solid line is the equivalence line. At the differences and a histogram of them are shown in
bottom and right side, respectively.} 
 \label{models}  
 \end{figure*}

In Fig.~ \ref{models} we show a comparison $\sim$3600 \w\ measured using \pacce\ and ` `hand-made" measurements with  {\sc iraf} 
{\it splot} task. We also plot the differences between the measurements and make a histogram of these differences to 
give a better idea of the dispersion between both sets. It is clear from Fig.~ \ref{models} that \pacce\ does 
reproduce very well the {\it splot} ``manual" \w\ values.  Note that there seems to be a very slight 
systematic under-prediction of the \w\ computed with \pacce\ if compared with those of {\it splot}, however, 
the differences are $\lesssim$0.05\AA. Similar results were obtained by \citet[][see their Fig. 6]{sousa07}, thus, suggesting that {\sc iraf} {\it splot} measurements may slightly overestimate the \w\ values.

\begin{figure*}[h]
\centering
\includegraphics[scale=0.80]{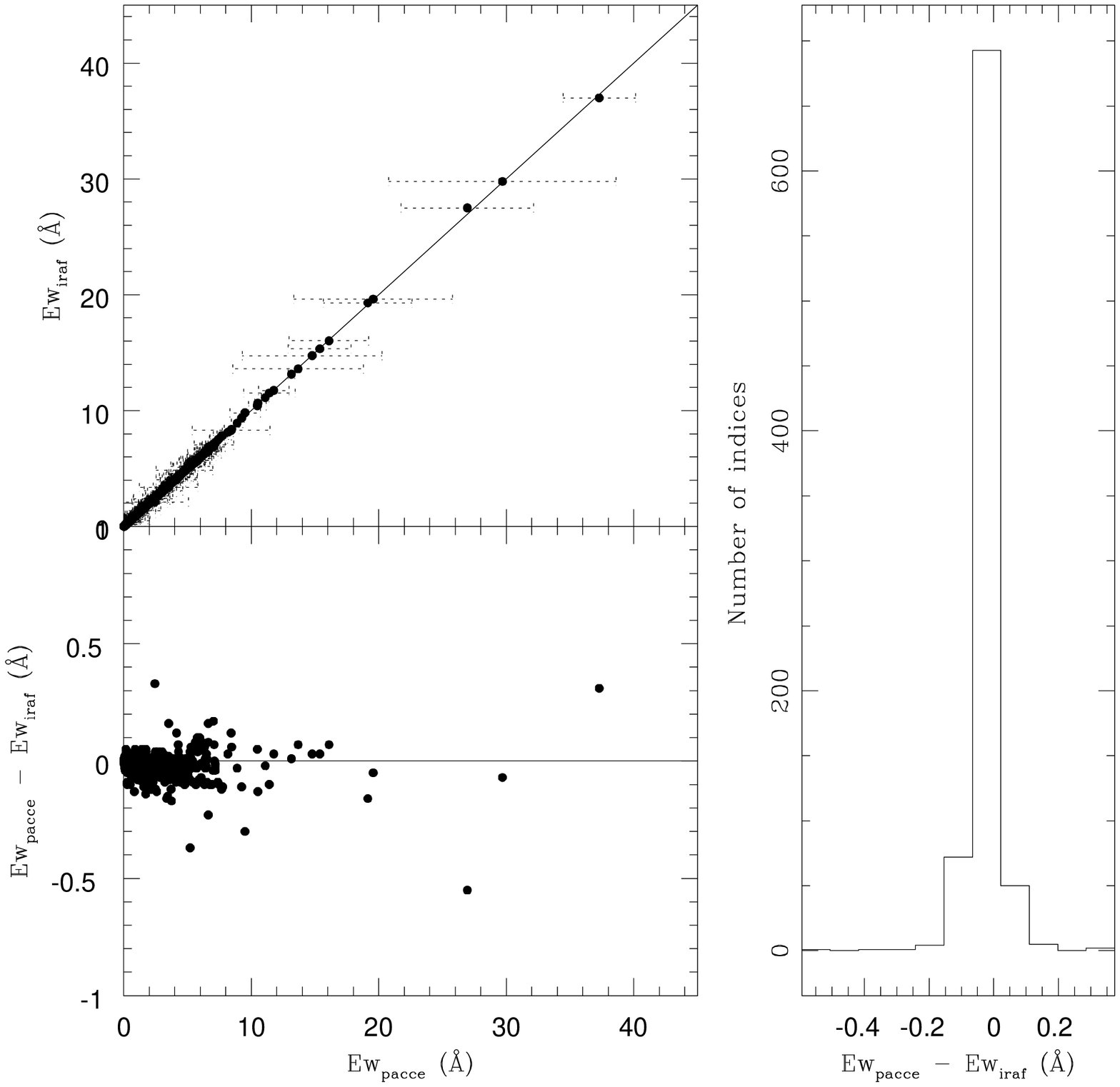}
 \caption{Same as Fig.~ \ref{models} but for observed optical stellar spectra. Error bars are shown.} 
 \label{data}  
 \end{figure*}

It is even harder to properly measure \w\ in observed spectra than in theoretical ones. To properly test our code 
capability in dealing with real spectra we have measured a set of \w\  ($\sim$ 950\w) in the
library of observed stellar spectra presented by \citet{silva92}. The results of 
such test are shown in Fig.~ \ref{data}. It is clear that \pacce\ is able to properly reproduce the 
measurements of {\it splot} task. However, a larger dispersion in the differences is observed
between both measurements than when using SSP models. In addition, these differences are clearly within the errors
and are lower than $\sim$0.5\AA.

\section{Final Remarks}\label{final}

We present \pacce\ a  Perl algorithm to compute continuum and \w.  We describe the method used in the computations, 
as well as the requirements for its use.

We compare the measurements made with \pacce\  and ``manual" ones made using {\sc iraf} {\it splot} task. These tests 
show that for SSP models (i.e. high S/N) the \w\ values are very similar (differences $\lesssim$\,0.2\,\AA). 
In real stellar spectra, the correlation between both values is also very good, but with differences of up to 0.5\,\AA.
However, these small differences can be explained by the intrinsic errors in subjectiveness determinations
of continuum levels caused in ``manual" measurements. 

\pacce\ is also able to determine mean continuum and continuum at line center values, which are helpful in stellar
population studies. In addition, it is also able to compute the $\sigma (W_{\lambda})$ using photon statistics.

\section*{Acknowledgments}

We thank an anonymous referee for helpful suggestions. We thank Miriani G. Pastoriza and Bas\'{i}lio X. Santiago for helpful discussions. TBV thanks Brazilian financial support agency CNPq.


\begin{thebibliography}{}

\bibitem[Bica(1988)]{bica88} Bica, E. \ 1988, A \& A, 195, 76.

\bibitem[Feltzing \& Gonzalez(2001)]{felt01} Feltzing, S. \& Gonzalez, G., 2001, A\&A, 367, 253.

\bibitem[Gonzalez \& Lambert(1996)]{gonz96} Gonzalez, G. \& Lambert, D. L. 1996, AJ, 111, 424.

\bibitem[Krabbe et al.(2007)]{krabbe07} Krabbe, A. C., Rembold, S. B. \& Pastoriza, M. G., 2007, 2007, MNRAS, 378, 569.

\bibitem[Krabbe et al.(2008)]{krabbe08} Krabbe, A. C.; Pastoriza, M. G.; Winge, C.; Rodrigues, I.; Ferreiro, D. L., 2008, MNRAS, 389, 1593.

\bibitem[Maraston(2005)]{maraston05} Maraston, C., 2005, MNRAS, 362, 799.


\bibitem[Pogge \& Owen(1993)]{pow93} Pogge, R. W., \& Owen, J. M. 1993, OSU Internal Report 93-01

\bibitem[Rembold \& Pastoriza(2007)]{rembold07} Rembold, S. B. \& Pastoriza, M. G., 2007, MNRAS, 374, 1056

\bibitem[Rickes et al.(2004)]{rickes04} Rickes, M. G., Pastoriza, M. G. \& Bonatto, Ch.,  2004, A\&A, 419, 449
\bibitem[Riffel et al.(2008)]{riffel08} Riffel, R.; Pastoriza, M. G.; Rodríguez-Ardila, A. \& Maraston, C., 2008, MNRAS, 388, 803.


\bibitem[Schmitt et al.(1996)]{schmitt96} Schmitt, H. R., Bica, E., \& Pastoriza, M. G.\ 1996, MNRAS, 278, 965 

\bibitem[Silva \& Cornell(1992)]{silva92} Silva, D. R.; Cornell, M. E.,  1992, ApJS, 81, 865.

	
\bibitem[Sousa et al.(2007)]{sousa07} Sousa, S. G., Santos, N. C., Israelian, G., Mayor, M. \& Monteiro, M. J. P. F. G., 2007, A\&A, 469, 783.

\bibitem[Vollmann, \& Eversberg(2006)]{volmann06} Vollmann, K. \& Eversberg, T., 2006, AN, 327, 862.

\bibitem[Worthey et al.(1994)]{worthey94} Worthey, G.; Faber, S. M.; Gonzalez, J. J. \& Burstein, D., 1994, ApJS, 94, 687.


\end{thebibliography}
\end{document}